\documentclass{elsart}
\usepackage{amsmath}
\usepackage{amssymb}

\begin{document}
\newtheorem{theorem}{Theorem}[section]
\newtheorem{lemma}[theorem]{Lemma}
\newtheorem{remark}[theorem]{Remark}
\newtheorem{definition}[theorem]{Definition}
\newtheorem{corollary}[theorem]{Corollary}
\def\deg{\operatorname{deg}}
\def\tr{{\operatorname{Tr}}}
\def\id{{\operatorname{id}}}
\def\Pspan{{\operatorname{Span}}}
\def\BB{{\mathcal{B}}}
\begin{frontmatter}
\title
{Duality symmetry of the $p$-form effective action and
super trace of the twisted de Rham
complex}
\author[MPI]{P. Gilkey}, \ead{gilkey@darkwing.uoregon.edu}
\author[MPI,Baylor]{K. Kirsten}, \ead{klaus.kirsten@mis.mpg.de}
\author[MPI,SPb]{D. Vassilevich}, \ead{\newline vassil@itp.uni-leipzig.de}
\author[Alberta,FIAN]{A. Zelnikov}\ead{\newline zelnikov@phys.ualberta.ca}
\address[MPI]{Max Planck Institute for 
Mathematics in the Sciences, 
Inselstrasse 22-26, 04103 Leipzig, Germany}
\address[Baylor]{Department of Mathematics, 
Baylor University, Waco, TX 76798, 
USA}
\address[SPb]{V.~Fock Institute of Physics, St.Petersburg University,
Russia}
\address[Alberta]{Theoretical Physics Institute, University of Alberta,
Edmonton Alberta T6G 2J1, Canada}
\address[FIAN]{Lebedev Physics Institute, Moscow, Russia}
\begin{abstract} 
We consider quantum $p$-form fields interacting with a background dilaton.
We calculate the variation with respect to the dilaton of a difference
of the effective actions in the models related by a duality transformation.
We show that this variation is defined essentially by
the supertrace of the twisted de Rham complex.
The supertrace is then evaluated on a manifold of an arbitrary dimension,
with or without boundary.
\end{abstract}
\end{frontmatter}
\section{Introduction}
Let $M$ be a smooth compact Riemannian manifold of dimension $m$. Let $A_p$
be a $p$-form field with the field strength $F=dA_p$. 
Let $\phi$ be a scalar field (dilaton). 
The classical action for this system
is then given by
\begin{equation}
S=\int_M e^{-2\phi}F\wedge \star F\,,\qquad F=d A_p \,.\label{SFA}
\end{equation}
Here $\star$ is the Hodge operator, $\star^2=(-1)^{p(m-p)}$.
Later we shall also use a normalized Hodge operator $\tilde\star$.
Interactions of this type appear in many physical applications.
Among them are various reductions from higher dimensions,
extended supergravities \cite{CSF78}, and bosonic
M theory \cite{HS00}. Instead of the dilaton, also a tachyon coupling
may appear \cite{KT98}.

We consider the case when the field $A_p$ is quantized while the dilaton
$\phi$ and the metric are kept as a classical background. 
The complete information about quantum properties of the system 
is contained in an effective action $W_p(\phi )$, which is a 
non-local functional of
background fields. All quantum mean values of physical observables 
and their correlators can be derived by variation of the effective
action with respect to $\phi$ and the background metric. 

The main aim of this paper is to
study the symmetry properties of $W_p(\phi )$ under the duality
transformation 
\begin{equation}
p\to m-p-2\,,\qquad \phi \to -\phi \,.\label{dualtr}
\end{equation} 
We shall demonstrate that the difference $W_p(\phi )-W_{m-p-2}(-\phi )$
is described by the supertrace of the twisted de Rham complex and calculate
the supertrace in any dimension $m$. We are motivated by higher 
dimensional supergravity theories where (\ref{dualtr}) is a part of the
$S$-duality transformation. 

Probably the first calculation of this type was performed when $m=2$ and $p=0$
by Schwarz and Tseytlin \cite{ST92} who used the methods of an earlier
paper \cite{Sch79}. A similar problem appears in 
two-dimensional dilaton gravity \cite{GKV02} where some symmetry
properties of the effective action for a scalar field coupled to the
dilaton were found in \cite{GZ99} by direct calculations. Also for
$m=2$, an exact expression for a class of fermionic determinants
was obtained \cite{KV98} using the supertrace of the Dirac complex.
A more systematic approach has been suggested in \cite{VZ00} where
the duality symmetry has been studied in dimension $2$ for a ``non-abelian''
(matrix-valued) dilaton field and for the dilaton--Maxwell theory
in dimension $4$.

As a historical side-remark we note that the twisted de Rham complex
has been also used in supersymmetric quantum mechanics \cite{ABI,Wit82} 
and in Morse theory \cite{Wit82}.

Here is a brief outline to this paper. Section \ref{sec2} contains some
basic definitions. In Section \ref{sec3}, we study the variation of the
effective action with respect to the dilaton. In particular, we demonstrate
that a variation of two effective actions in the models related by
the duality transformation (\ref{dualtr}) is essentially defined by a
specific combination of localized heat trace coefficients
(Theorem \ref{thm3.1}) which we identify with the supertrace. An explicit expression for
the supertrace
of the twisted de Rham complex is given by Theorem \ref{thm4.2}; the proof of this result
uses the associated functorial properties and techniques of invariance theory. In Section
\ref{Sect5}, we generalize these results to the category of manifolds with boundary. 
In Section \ref{sec8} we give some concluding remarks and calculate the
difference of two dual effective actions for $M=\mathbb{R}^m$.
\section{The twisted de Rham complex, heat trace, and zeta function}\label{sec2}

Let $(M,g)$ be a compact Riemannian manifold
without boundary. Let $\phi$ be an auxiliary smooth function on $M$
which we use to twist the exterior derivative operator $d$ by setting:
$$d_\phi:=e^{-\phi}de^\phi
\quad\text{on}\quad C^\infty(\Lambda M)\,.$$
Let $\delta_{\phi,g}$ and $\Delta_{\phi,g}$ be the associated twisted coderivative and twisted
Laplacian:
$$\delta_{\phi,g}:=e^\phi\delta_g e^{-\phi}\qquad\text{and}\qquad
  \Delta_{\phi,g}:=(d_\phi+\delta_{\phi,g})^2=d_\phi\delta_{\phi,g}+\delta_{\phi,g} d_\phi\,.$$
Since $d_\phi^2=0$, we have an elliptic complex
$d_\phi^p:C^\infty(\Lambda^p M)\rightarrow C^\infty(\Lambda^{p+1}M)$
and we use the $\mathbb{Z}$ grading of this complex to decompose:
$$\Delta_{\phi,g}=\oplus_p\Delta_{\phi,g}^p\quad\text{where}\quad
\Delta_{\phi,g}^p:C^\infty(\Lambda^p M)\rightarrow C^\infty (\Lambda^pM)$$
is a non-negative operator of Laplace type.
If we set $\phi=0$, then we recover the ordinary untwisted de Rham complex. Let
$$\chi(M):=\sum_p(-1)^p\dim H^p(M;\mathbb{R})$$
be the Euler-Poincar{\'e} characteristic of the manifold $M$. Since the
index of an elliptic complex is invariant under perturbations, we may use the
Hodge-de Rham isomorphism which identifies $\ker(\Delta_{0,g}^p)=H^p(M;\mathbb{R})$
to see that:
\begin{equation}
\operatorname{index}(d_\phi)=\operatorname{index}(d_0)
  =\sum_p(-1)^p\dim\ker(\Delta_{0,g}^p)
  =\chi(M).
\label{eqn1.b}\end{equation}
The Hodge decomposition theorem extends to the twisted setting. Thus there is an
orthogonal direct sum decomposition
$$C^\infty(\Lambda^pM)=d_\phi C^\infty(\Lambda^{p-1}M)\oplus
  \delta_{\phi,g} C^\infty(\Lambda^{p+1}M)\oplus\ker\Delta_{\phi,g}^p\,$$
which decomposes any form as the sum of a twisted exact, twisted co-exact, and twisted
harmonic form:
$$A_p=d_\phi A_{p-1}+\delta_{\phi,g}
A_{p+1}+\gamma_p\quad\text{for}\quad\gamma_p\in\ker\Delta_{\phi,g}^p.$$ The associated
projections on the spaces of twisted exact and co-exact forms will be denoted by the
subscripts $||$ and
$\perp$ respectively; note that the space of twisted harmonic forms
$\ker\Delta_{\phi,g}^p$ is finite dimensional.

 Let $D$ be an operator of Laplace
type on
$M$. The fundamental solution of the heat equation, $e^{-tD}$, is an infinitely smoothing
operator. Let
$f\in C^\infty(M)$ be an auxiliary smearing function. As
$t\downarrow0$, there is a complete asymptotic expansion:
\begin{equation}\tr_{L^2}(fe^{-tD})\sim\sum_{n=0}^\infty
t^{(n-m)/2}a_{n,m}(f,D)\label{asym}\end{equation}
where the {\it heat trace coefficients} $a_{n,m}(f,D)$ are locally computable; this means
that there are local invariants $a_{n,m}(D)(x)$ so that:
$$a_{n,m}(f,D)=\int_Mf(x)a_{n,m}(D)(x)dx$$
where $dx$ is the Riemannian element of volume.
The function $f$ localizes the question and permits us to recover the divergence
terms which would otherwise not be detected. The invariants
$a_{n,m}$ vanish identically if
$n$ is odd.

Suppose additionally that $D$ is non-negative. 
Let $\Pi_D$ be orthogonal projection
on the zero eigenspace and let
$\widehat D$ be $D$ acting on $\ker(D)^\perp$ (i.e. project out
the zero eigenspace). Define the smeared zeta function by setting:
$$
\zeta_D(f,s):=\tr_{L^2} (f\widehat D^{-s}) \,.
$$
The zeta function is
related to the heat trace by means of the Mellin transform
\begin{equation}
\zeta_D (f,s)=
\frac 1{\Gamma (s)} \int_0^\infty t^{s-1} \tr_{L^2} 
\{f(e^{-tD}-\Pi_D)\}dt\,.
\label{Mellin}
\end{equation}
Since $\tr_{L^2}\{f(e^{-tD}-\Pi_D)\}$ 
decays exponentially as $t\rightarrow\infty$, eqn.
(\ref{asym}) and eqn. (\ref{Mellin}) imply that the zeta function
$\zeta_D(f,s)$ is regular at $s=0$ and that
$$
\zeta_D(f,0)=a_{m,m}(f,D) -\tr_{L^2} \{ f \Pi_D\} \,.
$$
Let prime denote differentiation with respect 
to the parameter $s$. We use the zeta
function to define the determinant of the operator
$D$ by setting \cite{dowk76-13-3224,hawk77-55-133,ray73-98-154}:
\begin{equation}
\ln\det \widehat D:=-\zeta'_D(1,0) \,.\label{detz}
\end{equation}

\section{The effective action}\label{sec3}
We consider quantum effects in a system described by the action given in eqn.
(\ref{SFA}).
It is natural to include the dilaton field $\phi$ as well in the definition of
the inner product by setting:
\begin{equation}
\langle A_p,B_p \rangle =\int_M e^{-2\phi} A_p \wedge \star B_p
\,.\label{inpr}
\end{equation}
This equation defines also the path integral measure. For a different 
measure, the effective action will receive a contribution from the
scale anomaly. This contribution is relatively easy to control.
For the rest of this paper we adopt the measure defined by 
(\ref{inpr}). The fields $\tilde A_p:=e^{-\phi}A_p$ have a
standard Gaussian measure and are to be considered as fundamental
fields in the path integral. The action given in eqn. (\ref{SFA}) is invariant
under the gauge transformation which sends $\tilde A_p$ to $\tilde A_p +
d_\phi \tilde A_{p-1}$. This means that the $p$-forms which are
$d_\phi$ exact have to be excluded from the path integral, but that
a Jacobian factor corresponding to the ghost fields $\tilde A_{p-1}$
has to be included in the path integral measure. Next we note that
$d_\phi$-exact $(p-1)$-forms do not generate a non-trivial transformation
of $\tilde A_p$. Hence, such fields must be excluded from the ghost
sector. Then we have to include ``ghosts for ghosts''. This goes on
until the zero forms have been reached. By giving these arguments
an exact meaning, one arrives at the Faddeev--Popov approach to
quantization of the $p$-form actions \cite{Bar95,DPvN,Ob82,Si80}. 
We note that the procedure of \cite{Bar95,Ob82} is valid also in the
presence of a dilaton interaction if one simply replaces the ordinary
derivatives by the twisted ones. As a result, we have the following 
expression for the
effective action:
\begin{equation}
W_p(\phi ):=\frac 12 \sum_{k=0}^p (-1)^{p+k}
\ln \det (\Delta_{\phi,g}^k\vert_\perp )
+W_p^{\rm top}
\label{ea}
\end{equation}
where all determinants are restricted to the spaces of twisted co-exact
forms. Let $b_j:=\dim H^j(M;\mathbb{R})$ be the $j^{th}$ Betti number. According to
\cite{Bar95} the ``topological'' part of the effective action  is given
by\footnote{Slightly different quantization schemes yield different $W_p^{\rm top}$ (cf.
\cite{Bar95,DPvN}). Some choices of
$W_p^{\rm top}$ may lead to non-local contributions to the
scale anomaly \cite{ELV96}.
}: 
$$W_p^{\rm top}:= \ln \sigma \sum_{j=0}^p (-1)^{p+j} b_j.$$
The coupling constant $\sigma$ (an overall
factor in front of the action (\ref{SFA})) has not been written
explicitly. The topological effective action $W_p^{\rm top}$ will play no role in the
subsequent  calculations since it does not depend on $\phi$.

The twisted exterior derivative $d_\phi$ intertwines 
$\Delta_{\phi,g}^p\vert_\parallel$ and $\Delta_{\phi,g}^{p-1}\vert_\perp$.
This  shows for the total zeta function in terms of the coexact ones,
$$
\zeta_{\Delta_{\phi,g}^p} (1,s) = \zeta_{\Delta_{\phi , g} ^p} \vert_\perp 
(1,s) + \zeta_{\Delta_{\phi , g}^{p-1}} \vert_\perp (1,s) , 
$$
the inverse of which is
$$
\zeta_{\Delta_{\phi , g} ^p} \vert_\perp 
(1,s) = \sum_{k=0}^p (-1)^{p+k} \zeta_{\Delta_{\phi,g}^k} (1,s) .
$$
Each individual term in (\ref{ea}) can be calculated via (\ref{detz})
and we have
\begin{equation}
W_p(\phi )=-\frac 12 \sum_{k=0}^p (-1)^{p+k}  
\zeta^\prime_{\Delta_{\phi,g}^k}\vert_\perp (1,0)
+W_p^{\rm top}\,.\label{ea2}
\end{equation}
In general, dealing with the zeta functions, we use the method
of \cite{APS76,RS79}. We first assume that $s$ is sufficiently large
to keep us away from the singularities, then use the Mellin transformation described in
eqn. (\ref{Mellin}), perform the variation, then perform the Mellin transformation
backwards, and then continue the results to $s=0$. This is a perfectly
standard procedure which allows us to work with variations of positive
integer powers of $D$; see
\cite{esposito} for further details.

We consider the variation of $\zeta_{\Delta_{\phi,g}^p \vert_\perp}(1,s)$
under an infinitesimal variation of $\phi$ and remark that the symbol `$\delta$' has two
different meanings in the following equation:
\begin{eqnarray}
&&\delta \zeta_{\Delta_{\phi,g}^p \vert_\perp}(1,s)=
\delta \tr_{L^2} ((\delta_{\phi,g} d_\phi \vert_{\Lambda^p_\perp})^{-s})
=\delta \tr _{L^2}
((e^\phi \delta_ge^{-2\phi} d e^\phi \vert_{\Lambda^p_\perp})^{-s}) 
\nonumber \\
&&\quad = -2s 
\tr_{L^2} \left( (\delta \phi) \delta_{\phi,g} d_\phi  (\Delta_{\phi,g}^p)^{-s-1} 
\vert_{\Lambda^p_\perp} \right)\nonumber\\&&\qquad
+2s \tr_{L^2} \left( (\delta \phi) d_\phi \delta_{\phi,g}  (\Delta_{\phi,g}^p)^{-s-1} 
\vert_{\Lambda^{p+1}_\parallel} \right) \nonumber\\
&&\quad = -2s \left( \zeta_{\Delta_{\phi,g}^p\vert_\perp} (\delta\phi,s)
- \zeta_{\Delta_{\phi,g}^{p+1}\vert_\parallel} (\delta\phi,s) \right) \,.
\label{delze}
\end{eqnarray}
This formal derivation can be justified using the procedure outlined above.
We remark that the localized zeta function 
$ \zeta_{\Delta_{\phi,g}^p\vert_\parallel} (\delta\phi,s)$ may have a pole
at $s=0$ (for an explicit example see \cite{VZ00}) while 
$ \zeta_{\Delta_{\phi,g}^p\vert_\parallel} (1,s)$ and
$ \zeta_{\Delta_{\phi,g}^p} (\delta\phi,s)$ are regular at this point.

Our aim is to study symmetry properties of $W_p(\phi )$ with respect to the
reflection $\phi\to -\phi$. To proceed further we need several 
identities between spectral functions of the Laplacians with
$\phi$ and $-\phi$.  

Let $\tilde\star_g$ be the normalized Hodge operator defined by a local
orientation of $M$.
Then, the normalizations having taken into account the
sign conventions, one has the following relationships:
\begin{equation}\begin{array}{ll}
\tilde\star_g^2=\text{id},&
\tilde\star_g\phantom{.}d_\phi\phantom{.}\tilde\star_g=\delta_{\phi,g},\\
\tilde\star_g\phantom{.}\delta_{\phi,g}\phantom{.}\tilde\star_g=d_\phi,&
\tilde\star_g\phantom{.}\Delta_{\phi,g}^p\phantom{.}\tilde\star_g=
\Delta_{\phantom{.}-\phi,g}^{m-p},\\
\tilde\star_g\phantom{.}\Delta_{\phi,g}^p\vert_\perp \phantom{.}\tilde\star_g=
\Delta_{\phantom{.}-\phi,g}^{m-p}\vert_\parallel,\qquad\qquad&
\zeta_{\Delta_{\phi,g}^p\vert_\perp}(\delta \phi,s)=
\zeta_{\Delta_{-\phi,g}^{m-p}\vert_\parallel}(\delta\phi,s)\,.
\end{array}\label{zeze}
\end{equation}
\begin{theorem}\label{thm3.1} We have that
$$
\delta \left( W_p(\phi )-W_{m-p-2}(-\phi )\right)=
\sum_{k=0}^m (-1)^{p+k} \left[ a_{m,m}(\delta\phi,\Delta_{\phi,g}^k)
- \tr_{L^2} \{ (\delta \phi ) \Pi_{\Delta^k_{\phi, g}} \} \right] \, .
$$
\end{theorem}
{\it Proof.}
We first use eqn. (\ref{delze}) and eqn. (\ref{zeze}) to demonstrate that
\begin{eqnarray}
&&-\frac 12 \delta \left[ \sum_{k=0}^p (-1)^{k+p}
\zeta_{\Delta_{\phi,g}^k \vert_\perp}(1,s) -
\sum_{q=0}^{m-p-2} (-1)^{p+q}
\zeta_{\Delta_{-\phi,g}^q \vert_\perp}(1,s) \right]\nonumber\\
&&=
s\sum_{k=0}^m (-1)^{p+k} \zeta_{\Delta_{\phi,g}^k} (\delta\phi,s) \,.\label{eqn3.1xx}
\end{eqnarray}
We differentiate eq. (\ref{eqn3.1xx}) with respect to $s$, take
the limit $s\to 0$, and use eq. (\ref{ea2}) together with properties
of the zeta function described in Section \ref{sec2} to complete the proof.
The term with the harmonic forms takes care of the zero mode subtraction.
$\Box$

\section{Supertrace of the twisted de Rham complex} Motivated by the term appearing in
Theorem \ref{thm3.1}, we express
\begin{eqnarray}
\sum_p(-1)^p\tr_{L^2}(fe^{-t\Delta_{\phi,g}^p})&\sim&\sum_{n=0}^\infty
t^{(n-m)/2}\int_Ma_{n,m}^{d+\delta}(\phi,g)(x)f(x)dx\quad\text{where}\nonumber\\
a_{n,m}^{d+\delta}(\phi,g):&=&\sum_{p=0}^m(-1)^pa_{n,m}(\Delta_{\phi,g}^p)\,.
\label{eqn-Pengui-14}
\end{eqnarray}
The cancellation argument of McKean and Singer \cite{McSi67} together with eq.
(\ref{eqn1.b}) then yields, setting the smearing function $f=1$, that:
\begin{equation}\int_Ma_{n,m}^{d+\delta}(\phi,g)(x)dx
=\left\{\begin{array}{ll}
\chi(M)&\text{if }m=n,\\
0&\text{if }m\ne n.\end{array}\right.\label{eqn-Pengui-15}\end{equation}
From eq. (\ref{eqn-Pengui-15}) 
we immediately see that for constant $\delta\phi$ 
the variation in Theorem \ref{thm3.1} is zero. The same statement can be
also derived from the invariance of $d_\phi$ and $\delta_{\phi,g}$
under constant shifts of the dilaton.

Let $R_{ijkl}$ be the components of the Riemann curvature tensor relative to a
local orthonormal frame with the sign convention that $R_{1221}=+1$ on the standard
sphere in $\mathbb{R}^3$.  We define the anti-symmetric permutation tensor
$\varepsilon$ by setting:
$$\varepsilon^{i_1...i_m;j_1...j_m}:=(e^{i_1}\wedge...\wedge
e^{i_m},e^{j_1}\wedge...\wedge e^{j_m})\,.$$
We adopt the Einstein convention and sum
over repeated indices.  If $m=2\bar m$ is
even, then the {\it Pfaffian} or {\it Euler form} is defined by setting:
\begin{equation}\mathcal{E}_m:=(4\pi)^{-\bar m}\textstyle\frac1{2^{\bar m}\bar m!}
\varepsilon^{i_1...i_m;j_1...j_m}R_{i_1i_2j_2j_1}...
R_{i_{m-1}i_mj_{m}j_{m-1}}.
\label{eqn-Pengui-16}\end{equation}
We set $\mathcal{E}_m=0$ if $m$ is odd. For example, we have:
$$\textstyle\mathcal{E}_2=\frac1{4\pi}R_{ijji}\quad\text{and}\quad
\mathcal{E}_4
=\frac1{32\pi^2}((R_{ijji})^2-4|R_{ijjk}|^2+|R_{ijkl}|^2)\,.$$

The following theorem of Patodi \cite{Pa70} deals with the
untwisted case:
\begin{theorem}\label{thm4.1} If $\phi=0$, then $a_{n,m}^{d+\delta}=0$ if $n<m$ and 
$a_{m,m}^{d+\delta}=\mathcal{E}_m$.
\end{theorem}

Combining
Theorem
\ref{thm4.1} with eq. (\ref{eqn-Pengui-15}), yields a heat equation proof of the
Chern-Gauss-Bonnet theorem
\cite{C44} for manifolds without boundary:
$$\chi(M)=\int_M\mathcal{E}_m(g)(x)dx\,.$$

Shortly after Patodi's original proof, Atiyah,
Bott, and Patodi \cite{ABP73} and Gilkey \cite{PG73} also gave proofs of this
result using very different methods. Subsequently many proofs of the index theorem
using heat equation methods have been given; we refer to \cite{BGV91} for an excellent
historical survey.

In the present note, we shall follow the development in
\cite{PG73} to extend Theorem \ref{thm4.1} to the twisted setting:

\begin{theorem}\label{thm4.2} For arbitrary $\phi$, $a_{n,m}^{d+\delta}=0$ if $n<m$
and
$a_{m,m}^{d+\delta}=\mathcal{E}_m$.
\end{theorem}

Theorem \ref{thm4.2} shows that the crucial term where $n=m$ does {\bf not} involve
$\phi$. We remark that the higher divergence terms  
where $n>m$ do involve $\phi$
\cite{GKV02a}.

The proof of Theorem \ref{thm4.2} will rest upon a detailed analysis of the invariants
$a_{n,m}^{d+\delta}$. We begin by introducing some spaces of invariants.
Let
$\mathcal{Q}_{n,m}$ be the set of all invariants which are homogeneous
of weight $n$ in the derivatives of the metric and the derivatives of an auxiliary
function $\phi$ with coefficients which are smooth functions of the metric tensor
and which are defined on underlying manifolds of dimension $m$. These spaces are
trivial if $n$ is odd. Let $\mathcal{P}_{n,m}\subset\mathcal{Q}_{n,m}$ be
the subspace of invariants not involving the auxiliary function $\phi$. For example,
$\phi_{;ii}\in\mathcal{Q}_{2,m}$, 
$R_{ijji;kk}\in\mathcal{P}_{4,m}$, and
$\mathcal{E}_m\in\mathcal{P}_{m,m}$.

The following natural restriction map
$r:\mathcal{Q}_{n,m}\rightarrow\mathcal{Q}_{n,m-1}$
 will play a crucial role. If $(N,g_N,\phi_N)$ are
structures in dimension $m-1$, then we can define
corresponding structures in dimension $m$ by setting 
$$(M,\phi_M,g_M):=(N\times
S^1,\phi_N,g_N+d\theta^2)\,.$$
If
$x\in N$ is the point of evaluation, we take the corresponding point $(x,1)\in M$ for
evaluation; which point on the circle chosen is, of course, irrelevant as $S^1$ has a
rotational symmetry. The restriction map is then characterized dually by the formula:
$$r(Q)(\phi_N,g_N)(x)=Q(\phi_N,g_N+d\theta^2)(x,1)\,.$$
We can also describe the restriction map $r$ in classical terms. One considers polynomials
in the covariant derivatives of the function $\phi$ and of the curvature tensor $R$. The
theorem of H. Weyl
\cite{We46} on the invariants of the orthogonal group then 
shows that all invariants are
built by contracting indices in pairs, where the indices range from $1$ thru $m$. If
$P$ is given in terms of a Weyl spanning set, then $r(P)$ is given in terms of the same
Weyl spanning set by simply restricting the range of summation to be from $1$ thru $m-1$.

\begin{lemma}\label{lem4.3} We have $a_{n,m}^{d+\delta}\in\mathcal{Q}_{n,m}\cap\ker(r)$.
\end{lemma}

\noindent{\it Proof.} 
The heat trace invariants $a_{n,m}^{d+\delta}$ are homogeneous of degree $n$ in the jets of
the metric and of $\phi$. Thus $a_{n,m}^{d+\delta}\in\mathcal{Q}_{n,m}$.

If $D$ is any operator of Laplace type on a vector bundle $V$, then we have that
$a_0(x,D)=(4\pi)^{-m/2}\dim V$. Consequently, $a_{0,m}^{d+\delta}=0$
for any dimension $m$ as $\sum_p (-1)^p \dim(\Lambda^p M)=0$.
On the circle, the metric is flat. If
$\phi=0$, then the jets of $\phi$ play no role and thus $a_{n,1}^{d+\delta}=0$ for $n>0$.
This shows
\begin{equation} a_{n,1}^{d+\delta}(0,d\theta^2)=0\quad\text{for all}\quad n.
\label{eqn-Pengui-17}\end{equation}
Suppose the structures decouple. Then we may decompose
$\Lambda M=\Lambda M_1\otimes\Lambda M_2$, $d_\phi=d_1+d_2$ and
$\delta_{\phi,g}=\delta_1+\delta_2$ where on $C^\infty(\Lambda^pM_1\otimes\Lambda^qM_2)$ we
have:
$$\begin{array}{ll}
d_1:=d_{\phi_1}\otimes\id,&d_2:=(-1)^{p\phantom{.}}\id\otimes d_{\phi_2},\\
\delta_1:=\delta_{\phi_1,g}\otimes\id,&
\delta_2:=(-1)^{p\phantom{.}}\id\otimes\delta_{\phi_2,g_2}.
\end{array}
$$
Consequently these operators satisfy the commutation
relations:
$$d_1d_2+d_2d_1=0,\ d_1\delta_2+\delta_2d_1=0,\ \delta_1d_2+d_2\delta_1=0,\ 
\delta_1\delta_2+\delta_2\delta_1=0\,.$$
Thus we may express
$\Delta_{\phi,g}=\Delta_{\phi_1,g_1}\otimes\id+\id\otimes\Delta_{\phi_2,g_2}$ so the
fundamental solution of the heat equation is given by
$e^{-t\Delta_{\phi,g}}=e^{-t\Delta_{\phi_1,g_1}}\otimes e^{-t\Delta_{\phi_2,g_2}}$.
By taking the super trace and by equating
coefficients of $t$ in the resulting asymptotic expansions we see that:
\begin{equation}
a_{n,m}^{d+\delta}(\phi,g)(x_1,x_2)
=\sum_{n_1+n_2=n}a_{n_1,m_1}^{d+\delta}(\phi_1,g_1)(x_1)\cdot
a_{n_2,m_2}^{d+\delta}(\phi_2,g_2)(x_2).
\label{eqn-Pengui-18}
\end{equation}
We now set $(M_2,\phi_2,g_2)=(S^1,0,d\theta^2)$ and apply eqn. (\ref{eqn-Pengui-17}) and eqn.
(\ref{eqn-Pengui-18}) to see $r(a_{n,m}^{d+\delta})=0$.
$\Box$

The heat equation proof of the Chern-Gauss-Bonnet theorem given in \cite{PG73} relied
heavily on the following result:

\begin{lemma}\label{lem4.4} Let $P\in\mathcal{P}_{n,m}\cap\ker(r)$. If $n<m$, $P=0$.
If $n=m$, $P=c_m\mathcal{E}_m$.
\end{lemma}

We can extend this Lemma to the situation at hand by showing that $\phi$ plays no role for
$n\le m$:

\begin{lemma}\label{lem4.5} Let $Q\in\mathcal{Q}_{n,m}\cap\ker(r)$. If $n<m$, $Q=0$.
If $n=m$, $Q=c_m\mathcal{E}_m$.
\end{lemma}

{\it Proof.} Let $Q\in\mathcal{Q}_{n,m}\cap\ker(r)$. Fix a point $x_0\in M$ and
introduce a system of local coordinates
$x=(x^1,...,x^m)$ so that 
$$g_{ij}(x_0)=\delta_{ij}\quad\text{and}\quad
    \partial_kg_{ij}(x_0)=0\quad\text{ for }\quad
1\le i,j,k\le m\,.$$
For example,
geodesic polar coordinates centered at $x_0$ have this property. If
$\alpha:=(a_1,...,a_m)$ and $\beta:=(b_1,...,b_m)$ are multi-indices, introduce
variables
$$\phi_{/\alpha}:=\partial_\alpha^x\phi\qquad\text{and}\qquad
g_{ij/\beta}:=\partial_\beta^xg_{ij}$$
for the ordinary partial (not covariant) derivatives of the function $\phi$ and
of the metric $g$. We emphasize, we are {\bf not} working invariantly. With the
normalizations given above, we may then express:
\begin{equation}Q=Q(\phi_{/\alpha},g_{ij/\beta})\qquad\text{for}\qquad|\alpha|\ge1
\quad\text{and}\quad|\beta|\ge2\,.\label{eqn-Pengui-19}\end{equation}
Suppose that $Q\ne0$ and that $r(Q)=0$. Then $Q$ vanishes on product manifolds
$N\times S^1$ where the structures are flat in the $S^1$ direction. Restricting to
such product structures simply imposes the additional relations 
$$\phi_{/\alpha}=0\quad\text{and}\quad
g_{ij/\beta}=0\qquad\text{if}\qquad
a_m>0\quad\text{and}\quad \delta_{i,m}+\delta_{j,m}+b_m>0\,.$$
We let $\deg_\ell$ be the number of times the index $\ell$ appears in a given
variable:
$$\deg_\ell(\phi_{/\alpha}):=a_\ell\qquad\text{and}\qquad
\deg_\ell(g_{ij/\beta}):=\delta_{i\ell}+\delta_{j\ell}+b_\ell.$$
The
condition $r(Q)=0$ is then equivalent to the condition that $\deg_m(A)>0$ for every
monomial $A$ of $Q$ and, by symmetry, that $\deg_i(A)>0$ for every
monomial $A$ of $Q$ and every index $i$. Let
$$A=\phi_{/\alpha_1}...\phi_{/\alpha_u}g_{i_1j_1/\beta_1}...g_{i_vj_v/\beta_v}$$
be a monomial of $Q$. Since $Q$ is invariant under the change of coordinates
$x_m\rightarrow-x_m$ and $x_i\rightarrow x_i$ for $i<m$, the index $m$ (and hence
every index) must appear an even number of
times in
$A$. Consequently as every index appears in $A$, every index appears at least twice.
Using eqn. (\ref{eqn-Pengui-19}), we count indices to estimate:
\begin{equation}\begin{array}{l}
2m\le\sum_i\deg_i(A)
=\sum_\mu|\alpha_\mu|+\sum_\nu(2+|\beta_\nu|)=n+2v
\\
\le n+\sum_\nu|\beta_\nu|=2n-\sum_\mu|\alpha_\mu|
\le 2n-u\le 2n.\end{array}\label{eqn-Pengui-20}\end{equation}
This is impossible if $n<m$ so $Q=0$ if $n<m$.
On the other hand, if $n=m$, all of the inequalities in eqn. (\ref{eqn-Pengui-20}) must
have been equalities. Thus,
$u=0$ so $Q$ does not involve $\phi$ at all. Consequently
$Q\in\mathcal{P}_{m,m}\cap\ker(r)$ so by Lemma
\ref{lem4.4}, $Q=c_m\mathcal{E}_m$.
$\Box$

{\it Proof of Theorem \ref{thm4.1}.} We apply Lemmas \ref{lem4.3} and
\ref{lem4.5}; assertion (1) is now immediate. If $m$ is odd, then
$a_{m,m}^{d+\delta}=0$ so there is nothing to prove. If
$m$ is even, then $a_{m,m}^{d+\delta}=c_m\mathcal{E}_m$ for some universal
constant
$c_m$. Evaluating the integral on $S^2\times...\times S^2$ with $f=1$ where $S^2$ is given
the standard metric and using eq. (\ref{eqn-Pengui-15}) then yields
$$
c_m\int_{S^2\times...\times
S^2}\mathcal{E}_mdx=\chi(S^2\times...\times S^2)=2^{\bar m}.
$$
Taking into effect the multiplicative nature of the constants and the factor of
$\frac1{\bar m!}$ we see that
$$c_m\int_{S^2\times...\times S^2}\mathcal{E}_mdx
=c_m\left\{\int_{S^2}\mathcal{E}_2dx_2\right\}^{\bar m}
=c_m\left\{\frac1{8\pi}\int_{S^2}4R_{1221}dx\right\}^{\bar m}=c_m2^{\bar m}.
$$
We combine these two equations to see $c_m=1$.
$\Box$

\section{Manifolds with boundary}\label{Sect5}

We now suppose that $M$ has a non-empty, smooth boundary $\partial M$. Let $D_\BB$ be
the realization of an operator $D$ of Laplace type with respect to suitable
boundary conditions defined by a local boundary operator $\BB$ as discussed, for example,
in \cite{BG90}. If
$f\in C^\infty(M)$, let
$\nabla_m^k(f)$ denote the
$k^{th}$ normal covariant derivative of $f$ on the boundary with respect to the inward
unit normal. There is a complete asymptotic expansion
$$\tr_{L^2}(fe^{-tD_\BB})\textstyle\sim\sum_{n\ge0}t^{(n-m)/2}a_{n,m}(f,D,\BB)$$
where the heat trace asymptotics $a_{n,m}$ are locally computable:
$$a_{n,m}(f,D,\BB)=\int_Mf(x)a_{n,m}(D)(x)dx+\sum_k\int_{\partial M}
\nabla_m^kf(y)\cdot a_{n,m,k}^{\partial M}(D,\BB)(y)dy$$
where $dy$ is the Riemannian volume element on $\partial M$.
The new feature
here is the presence of the normal derivatives of the smearing function $f$; the heat
kernel behaves asymptotically like a distribution near the boundary and these additional
terms reflect this behaviour. Furthermore, note that $a_{n,m}(f,D,\BB)$ can be
non-zero for
$n$ odd. We refer to \cite{BG90} for a further discussion of these matters.

Absolute and relative boundary conditions, which are motivated by index theory, may be
defined as follows; we refer to
\cite{PG94} for further details. Near the boundary, we normalize the choice of
coordinates
$x=(y,x^m)$ so that
$\partial_m$ is the inward geodesic normal and so that $y=(y^1,...,y^{m-1})$ are
coordinates on the boundary. Let
$I=\{1\le a_1<...<a_p\le m-1\}$  be a multi-index and let
$dy^I:=dy^{a_1}\wedge...\wedge dy^{a_p}$ be the corresponding tangential
differential form on
$\partial M$. We decompose an arbitrary differential form $\omega$ into tangential and
normal components:
$$\omega:=\theta_Idy^I+\psi_Jdy^J\wedge dx^m$$
and define the absolute boundary operator by setting:
$$\mathcal{B}_a\omega:=\{(\partial_m\theta_I)dy^I|_{\partial M}\}\oplus
\{\psi_Jdy^J|_{\partial M}\}\,.$$
Let $\Delta_{\phi,g,\BB_a}$ be the realization of the twisted
Laplacian on the domain $\ker\mathcal{B}_a$. As before,
let $\tilde \star _g$ be the normalized Hodge operator.
Relative boundary conditions are given
by setting $\mathcal{B}_r:=\mathcal{B}_a\phantom{.}\tilde\star_g$ 
and the associated
realization of the Laplacian is denoted by $\Delta_{\phi,g,\BB_r}$. 
We may decompose
$$\Delta_{\phi,g,\BB_a}=\oplus_p\Delta_{\phi,g,\BB_a}^p
  \quad\text{and}\quad\Delta_{\phi,g,\BB_r}=\oplus_p\Delta_{\phi,g,\BB_r}^p\,.$$
We define the super-trace for absolute ($\mathcal{B}=\mathcal{B}_a$) or relative
($\mathcal{B}=\mathcal{B}_r$) boundary conditions by setting:
$$a_{n,m}^{d+\delta}(f,\phi,g,\mathcal{B}):=
\sum_p(-1)^pa_{n,m}(f,\Delta_{\phi,g}^p,\mathcal{B}).$$
This can then be expressed in terms of a local formula
$$a_{n,m}^{d+\delta}(f,\phi,g,\mathcal{B})=\int_Ma_{n,m}^{d+\delta}(\phi,g)(x)f(x)dx
+\sum_k\int_{\partial M}\nabla_m^kf(y)\cdot
a_{n,m,k}^{d+\delta,\mathcal{B}}(\phi,g)(y)dy$$
where the interior invariants $a_{n,m}^{d+\delta}$ are given by eqn. (\ref{eqn-Pengui-14})
and the boundary invariants are defined by
$$a_{n,m,k}^{d+\delta,\BB}(\phi,g)(y):=\sum_{p=0}^m(-1)^p
a_{n,m,k}(\Delta_{\phi,g}^p,\mathcal{B})(y).$$

If $\phi=0$, we have:
$$\begin{array}{ll}
\ker(\Delta_{g,a}^p)=H^p(M;\mathbb{R}),&\operatorname{index}(d_{\BB_a})=\chi(M)\\
\ker(\Delta_{g,r}^p)=H^p(M,\partial M;\mathbb{R}),\quad&
\operatorname{index}(\delta_{\BB_r})=\chi(M,\partial M).\end{array}
$$
If we set the smearing function $f=1$ and if $\phi$ satisfies Neumann boundary conditions,
then we recover the index:
$$
\begin{array}{l}
a_{n,m}^{d+\delta,\BB_a}(1,\phi,g)=\left\{\begin{array}{ll}
\chi(M)\phantom{,\partial M}&\text{if }m=n,\\
0&\text{if }m\ne n,\end{array}\right.\\
a_{n,m}^{d+\delta,\mathcal{B}_r}(1,\phi,g)=\left\{\begin{array}{ll}
\chi(M,\partial M)&\text{if }m=n,\\
0&\text{if }m\ne n.\end{array}\right.\\
\end{array}$$
This motivates the study of absolute and relative boundary conditions.

We return to the general setting and do not assume $\phi$ satisfies Neumann boundary
conditions. The normalized Hodge operator $\tilde\star_g$ intertwines relative and absolute
boundary conditions and intertwines $\Delta_{\phi,g}^p$ and $\Delta_{-\phi,g}^{m-p}$ so:
$$a_{n,m}^{d+\delta,\mathcal{B}_a}(f,\phi,g)
=(-1)^ma_{n,m}^{d+\delta,\mathcal{B}_r}(f,-\phi,g).$$
Consequently, we shall restrict to absolute boundary conditions henceforth.

If $\phi=0$, then these invariants were analyzed in \cite{PG75} where a proof of the
Chern-Gauss-Bonnet theorem \cite{C45} for manifolds with boundary was given using
heat equation methods. To describe those results, we must introduce some additional
invariants. Let $L$ be the second fundamental form of $\partial M\subset M$. Let
$m=2\bar m$ or $m=2\bar m+1$. Let $\{e_1,...,e_m\}$ be a local orthonormal frame near
the boundary where
$e_m$ 
is the inward geodesic normal. Let indices $a$, $b$ range from $1$ thru $m-1$.
For
$2k\le m-1$, define:
\begin{eqnarray*}
&&\mathcal{E}_{k,m}:=(8\pi)^{-k}\textstyle\frac1{k!(m-1-2k)!}\frac1{
\text{vol}(S^{m-1-2k})}
\varepsilon^{a_1,...,a_{m-1};b_1,...,b_{m-1}}\\&&\qquad\qquad\cdot
 R_{a_1a_2b_2b_1}...R_{a_{2k-1}a_{2k}b_{2k}b_{2k-1}}
L_{a_{2k+1}b_{2k+1}}...L_{a_{m-1}b_{m-1}}.
\end{eqnarray*}
In this definition, there are no $L$ terms if $2k=m-1$; there are no $R$ terms if
$k=0$.

We refer to \cite{PG75} for a proof of the following result:

\begin{theorem}\label{thm5.1} Let $\phi=0$. If $n<m-1$, then
$a_{n,m,0}^{d+\delta,\mathcal{B}_a}=0$. Furthermore,
$a_{m,m,0}^{d+\delta,\mathcal{B}_a}=\sum_k \mathcal{E}_{k,m}$.
\end{theorem}

Combining this result with the observations made above then gives a heat equation proof of
the Chern-Gauss-Bonnet theorem for manifolds with boundary.

We can generalize Theorem \ref{thm5.1} to this setting 
by removing the hypothesis that
$\phi=0$ and by showing the normal jets $f$ do not into the
crucial term when $n=m$:

\begin{theorem}\label{thm5.2}If $n+k\leq m-1$, then
$a_{n,m,k}^{d+\delta,\mathcal{B}_a}=0$. Furthermore,\newline
$a_{m,m,0}^{d+\delta,\mathcal{B}_a}=\sum_k \mathcal{E}_{k,m}$.
\end{theorem}

We illustrate Theorem
\ref{thm5.2} in dimensions $m=2,3,4$:
\begin{eqnarray*}
&&a_{2,2}^{d+\delta,\mathcal{B}_a}(f,\phi,g)=
    \textstyle\frac1{4\pi}\int_{M^2}fR_{ijji}
dx +\frac1{2\pi}\int_{\partial M^2}fL_{aa} dy,\\
&&a_{3,3}^{d+\delta,\mathcal{B}_a}(f,\phi,g)
   =\textstyle\frac1{8\pi}\int_{\partial
M^3}f\{ R_{a_1a_2a_2a_1}+L_{a_1a_1}L_{a_2a_2}-L_{a_1a_2}L_{a_1a_2}\}dy\\
&&a_{4,4}^{d+\delta,\mathcal{B}_a}(f,\phi,g)=
\textstyle\frac1{32\pi^2}\int_{M^4}f((R_{ijji})^2
-4|R_{ijjk}|^2+|R_{ijkl}|^2)dx\\
&&\textstyle\qquad\qquad\qquad
+\textstyle\frac1{24\pi^2}\int_{\partial M^4} f
\{ 3R_{abba}L_{cc}+6R_{acbc}L_{ab}+2L_{aa}L_{bb}L_{cc}\\
&&\textstyle\qquad\qquad\qquad
-6L_{ab}L_{ab}L_{cc}+4L_{ab}L_{bc}L_{ac}\}dy.
\end{eqnarray*}

Let
$\mathcal{Q}_{n,m}^{\partial M}$ be the set of all boundary invariants which are
homogeneous of weight $n$ in the derivatives of the metric and the derivatives of an
auxiliary function $\phi$ with coefficients which are smooth functions of the metric tensor
and which are defined on underlying manifolds of dimension $m$ and let
$\mathcal{P}_{n,m}^{\partial M}\subset\mathcal{Q}_{n,m}^{\partial M}$ be the subspace of
boundary invariants not involving the auxiliary function $\phi$. These spaces are non
trivial for all $(n,m)$.

As before, taking a product with the circle where 
the structures are trivial on the
circle defines a dual restriction 
$r:\mathcal{Q}_{n,m}^{\partial M}\rightarrow
    \mathcal{Q}_{n,m-1}^{\partial M}$. The same argument as that given for manifolds
without boundary then shows
$a_{n,m,k}^{d+\delta,\mathcal{B}_a}\in\mathcal{Q}_{n-k-1,m}^{\partial M}$. Thus Theorem
\ref{thm5.2} will follow from the following result in invariance theory and from Theorem
\ref{thm5.1}.

\begin{lemma}\label{lem7.4} Let $Q\in\mathcal{Q}_{n,m}^{\partial M}\cap\ker(r)$. If
$n<m-1$, then $Q=0$. If $n=m-1$, then $Q\in
\mathcal{P}_{m-1,m}^{\partial M}$.
\end{lemma}

{\it Proof.} We normalize the coordinates on the boundary so that
$$g_{mm}=1,\quad g_{am}=0,\quad\text{and}\quad g_{ab/c}(x_0)=0\,.$$
Thus $Q$ is a polynomial in the variables
$$\{\phi_{/\alpha},g_{ij/\beta},g_{ab/m}\}
\quad\text{for}\quad|\alpha|\ge1\quad\text{and}\quad|\beta|\ge2\,.$$
We consider a monomial
$A=\phi_{/\alpha_1}...\phi_{/\alpha_u}g_{i_1j_1/\beta_1}...g_{i_vj_v/\beta_v}
g_{a_1b_1/m}...g_{a_wb_w/m}$.
We argue as before to see that $\deg_aA$ is even and non-zero for $1\le a\le m-1$.
Consequently, we may estimate:
\begin{eqnarray*}
2(m-1)&\le&\sum_a\deg_a(A)
\le\sum_\mu|\alpha_\mu|+\sum_\nu(2+|\beta_\mu|)+2w
   =2v+w+n\\
&\le&\sum_\nu|\beta_\mu|+w+n=2n-\sum_\mu|\alpha_\mu|
\le2n-u\le 2n.
\end{eqnarray*}
This implies that $Q=0$ if $n<m-1$, while if $n=m-1$, all the inequalities must have
been equalities. Thus
$u=0$ so
$\phi$ does not enter.
$\Box$

\section{Conclusions}\label{sec8}
In this paper we have analysed the variation with respect to the dilaton
of a difference of two effective actions in the models related by a
duality transformation. This variation is reduced to a contribution
from twisted harmonic forms and to a combination of the heat trace
coefficients (supertrace of the twisted de Rham complex). Direct
evaluation of these heat trace coefficients is possible in low 
dimensions only. By using functorial properties of the supertrace
we are able nevertheless to obtain an explicit formula in any
dimension, also on manifolds with boundaries.

Theorem \ref{thm3.1} defines the variation of the effective actions with
respect to the dilaton. It can be integrated to give a relation
between the dual actions. Since we do not have a closed expression
for the twisted harmonic forms we suppose $M=\mathbb{R}^m$ and impose a
fall-off condition on the dilaton field. 
We also assume that the metric is asymptotically flat so that
the Laplace operators $\Delta^p_{\phi,g}$ have no normalizable
zero modes (non-normalizable zero modes never appear in the path integral
and, therefore, must not be subtracted).
Also in this case 
$W_p(0 )=W_{m-p-2}(0)$ \cite{Bar95}\footnote{In $m=4$
this has been demonstrated by Fradkin and Tseytlin \cite{FT84}.
}. Let $\mathcal{E}_m$ be given by (\ref{eqn-Pengui-16}). We have:
$$
W_p(\phi )-W_{m-p-2} (-\phi )= (-1)^p \int_M \phi \mathcal{E}_m\,dx \, .
$$

One can extend the results presented in this paper also to domain
wall and brane-world geometries. The heat trace asymptotics for these
cases can be found in \cite{GKV01}.

\section*{Acknowledgements}
The research of P.G.\ was partially supported by the NSF (USA)
and the MPI (Leipzig), K.K.\ and D.V.\ were supported by the MPI
(Leipzig), and A. Z.  was supported by the Killam Trust. 

\end{document}